\documentclass[review]{elsarticle}

\usepackage{lineno,hyperref}
\modulolinenumbers[5]

\journal{Chaos Solitons and Fractals}
\usepackage[latin9]{inputenc}
\usepackage{amsmath}
\usepackage{amssymb}
\usepackage{graphicx}
\usepackage{color}
\makeatletter

\pdfpageheight\paperheight
\pdfpagewidth\paperwidth

\@ifundefined{date}{}{\date{}}
\makeatother


\begin{document}

\begin{frontmatter}

\title{Anomalous transport and observable average in the standard map}

\author[mymainaddress1]{Lydia Bouchara}

\author[mymainaddress1]{Ouerdia Ourrad}

\author[mymainaddress2]{Sandro Vaienti}

\author[mymainaddress2,mymainaddress3]{Xavier Leoncini\corref{mycorrespondingauthor}}
\cortext[mycorrespondingauthor]{Corresponding author}
\ead{xavier.leoncini@cpt.univ-mrs.fr}

\address[mymainaddress1]{Laboratory of Theoretical Physics, Faculty of Exact sciences, University
of Bejaia, 06000, Bejaia, Algeria}

\address[mymainaddress2]{Aix Marseille Université, Université de Toulon, CNRS, CPT UMR 7332,
13288 Marseille, France }

\address[mymainaddress3]{Center for Nonlinear Theory and Applications, Shenyang Aerospace
University, Shenyang 110136, China}

\begin{abstract}
The distribution of finite time observable averages and transport
in low dimensional Hamiltonian systems is studied. Finite time observable average distributions are computed,
from which an exponent $\alpha$ characteristic of how the maximum of the distributions scales
with time is extracted. To link this exponent to transport properties,  the characteristic exponent
$\mu(q)$ of the time evolution of the different moments of order $q$ related to transport are computed. 
As a testbed for our study the standard map is used. The stochasticity parameter $K$
is chosen so that either phase space is mixed with a chaotic sea and islands of stability or with only a chaotic sea. Our observations lead to a proposition of a law relating the slope in $q=0$ of the function $\mu(q)$ with the exponent $\alpha$.
\end{abstract}

\end{frontmatter}

\linenumbers

\section{Introduction}

The question of transport in Hamiltonian systems is a long standing
issue as it can be inferred from the vast literature on the matter
and references therein \cite{Chirikov79,Mackay84,RomKedar90b,Dana85,Zaslavsky97,Venegeroles09,Zaslavbook2005}
and the large domain of applications ranging from hot magnetized plasmas
to astronomy, chaotic advection, underwater acoustics etc... Beyond
the fully chaotic situation in which we usually can apply the central
limit theorem, and therefore still have a random walker picture in
mind, problems are still not clear when the phase space is mixed.

Indeed in this situation the system is not ergodic, in the sense that there is
not only one unique ergodic component, but instead there are regions
with chaos, and regions with regular motion.
When considering one-and half degrees of freedom system, one usually talks about a
picture with a stochastic sea and islands/regions of regular motion.
This coexistence can lead to some problems especially since it is
possible for Hamiltonian systems to have so called sticky islands.
This paper inscribes itself in this series and tries to tackle the
problem of transport using distributions of finite time observable
averages and their evolution as the average is computed over larger
and larger times. A first attempt using this approach was performed
in \cite{Leoncini08}, using a perturbed pendulum as a case study,
and it found its roots in the study of advection described in \cite{LKZ01},
where finite time averages were used to detect sticky parts of trajectories.
In the present case we simply use the standard  or Taylor-Chirikov
map \cite{Chirikov79}. This choice was motivated by the fact that
such maps can be directly computed from the flow of the so-called
kicked rotor, and, being a map, it allows us to perform fast numerical
simulations and gather enough data to have somewhat reliable statistics. 
The purpose of this paper is not a thorough study of transport in the standard map,
but to use this map as a testbed for our analysis of finite time observable averages.

Regarding the problem of transport in Hamiltonian systems, the standard
map has become over the years a classical case study. One of its advantages
is that it depends on just one control parameter $K$ and many attempts
were made to find the link between $K$ and a diffusion coefficient
\cite{Chirikov79,Dana85,Lichtenberg92,ZaslavBook98}. Depending on
the values of $K$, we can get a system which is very close to an
integrable one or one that is fully chaotic, with, in between, the picture
of a mixed phase space with a chaotic sea and regular islands. In
this last setting, we can face the so called stickiness phenomenon:
particle's trajectory originating from the chaotic sea can stay (stick)
for arbitrary large times in the vicinity of a stable region. This
type of phenomenon is able to generate long memory effects, which,
in turn, can generate so called anomalous transport also called anomalous
diffusion. In contrast with normal diffusion, the dynamics leads to
transport properties which can be far from the Gaussian-like processes,
and the second moment grows nonlinearly in time.

In the following section, we briefly introduce the standard map and
present the phase space and first results with the choice of parameters
we considered. Then we discuss and consider transport properties in
each system. We present the method and compute 
characteristic transport exponents. We confirm the multi-fractal nature
of transport in both two considered cases where the phase space is mixed, while transport
appears as diffusive in the global chaotic case. Finally, we investigate
the relation between $\alpha$, the characteristic exponent of the
evolution of the maximum of the distribution of finite-time observable
averages, and $\mu$, the characteristic exponent of the second moment
of transport associated to the observable. In \cite{Leoncini08},
a simple law was proposed, namely $\alpha=1-\mu/2$, our findings
lead to good agreements for two out of the three cases. As a consequence,
a slightly more general law is then proposed which captures all features;
and then we conclude.

\section{The standard map}

Before moving to more details, we remind the reader that the standard
map arises naturally as a Poincaré mapping of the kicked rotor model,
whose Hamiltonian writes
\begin{equation}
H=\frac{p^{2}}{2}-\omega_{0}^{2}\cos(q)\times\sum_{n=-\infty}^{\infty}\delta(t-n\tau)\:,\label{eq:Ham_kicked_rotor}
\end{equation}
where the parameters  $\omega_{0}$ and $\tau$ are without dimensionality and $\delta$ is the Dirac function.
We shall not derive the standard map here, and we will consider it on the torus. In this case
its equations are 
\begin{equation}
\left\{ \begin{array}{ll}
p_{n+1}=p_{n}+K\sin q_{n}\:[2\pi]\\
q_{n+1}=q_{n}+p_{n+1}\:[2\pi],
\end{array}\right.\label{eq:Standard map}
\end{equation}
\begin{figure}
\begin{centering}
\includegraphics[width=8cm]{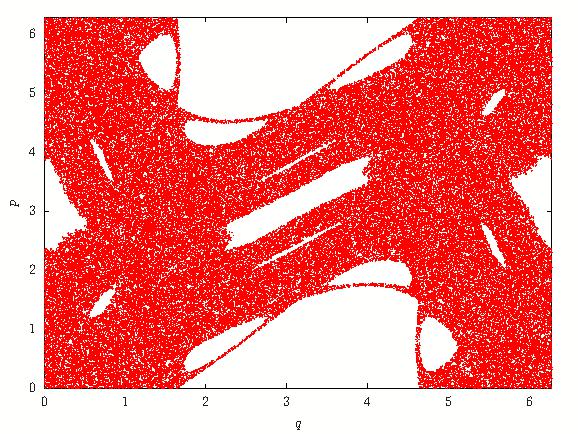}
\par\end{centering}

\protect\caption{Phase space visualization for $k=1.5$. We observe a mixed phase space
and Hamiltonian chaos which covers about half of the phase space.\label{fig:Phase-space-visualsationk=00003D1.5}}
\end{figure}
 
\begin{figure}
\begin{centering}
\includegraphics[width=8cm]{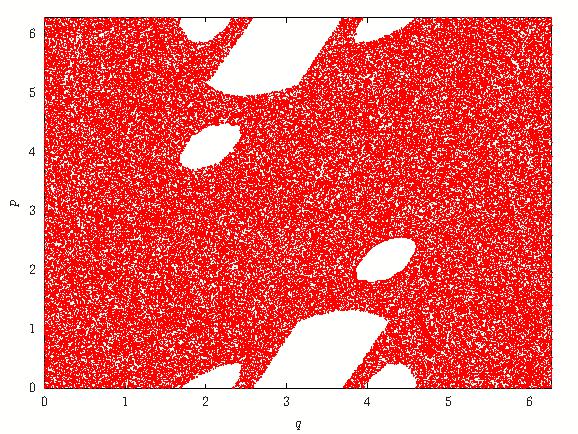}
\par\end{centering}

\protect\caption{Phase space visualization for $k=2.5$. We observe a mixed phase space
and Hamiltonian chaos which covers a larger portion than in Fig.~\ref{fig:Phase-space-visualsationk=00003D1.5}.\label{fig:Phase-space-visualsationk=00003D2.5}}
\end{figure}
\begin{figure}
\begin{centering}
\includegraphics[width=8cm]{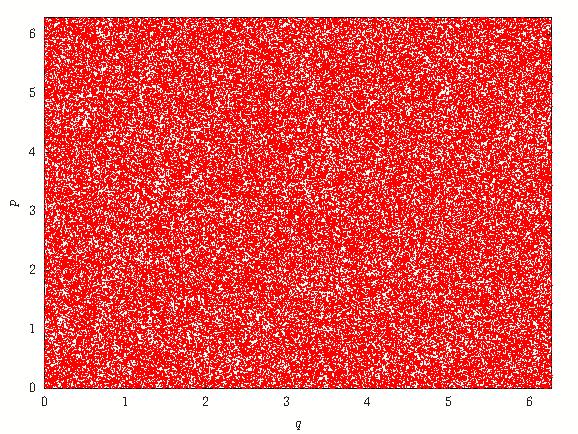}
\par\end{centering}

\protect\caption{Phase space visualization for $k=10$. We observe a fully chaotic
phase space and no regular islands subsist.\label{fig:Phase-space-visualsationk=00003D10}}
\end{figure}
where $K$ is the parameter that characterizes the force amplitude \cite{Chirikov79}.
Before going to study transport in this system, we shall briefly present
the three different cases considered. Namely, we considered three different
values for $K$. The plots for $K=1.5$, $K=2.5$, and $K=10$ are
represented respectively in Figs. \ref{fig:Phase-space-visualsationk=00003D1.5},\ref{fig:Phase-space-visualsationk=00003D2.5},
and \ref{fig:Phase-space-visualsationk=00003D10}.

Now that we specified the object of our study, let us consider transport.

\section{Transport Properties}

In order to consider transport, we shall first consider an observable.
In previous studies \cite{Leoncini08,LKZ01,LZ02,Leoncini05}, it has
been found that considering the absolute speed as an observable, leads
to relatively clear results for transport. However in the previously
mentioned works, the considered Hamiltonian systems were flows, and the study of
transport was made by considering the dispersion of the arc-length
of different trajectories. Here we  consider the standard map (\ref{eq:Standard map}).
The notion of arc-length does not make much sense unless we consider
the underlying kicked-rotor flow. Though, we may consider an analog
to the norm of the phase space speed, by considering the distance
between the two points $(q_{n+1},p_{n+1})$ and $(q_{n},p_{n})$;
given the equations (\ref{eq:Standard map}). This unfortunately does
not define a proper observable since it mixes both steps $n$ and
$n+1$. In order to get a real phase space observable we define one
inspired from this distance by 
\begin{equation}
v=\sqrt{K^{2}\sin^{2}q+p^{2}}\:.\label{eq:speed-1}
\end{equation}
\begin{figure}
\begin{centering}
\includegraphics[width=8cm]{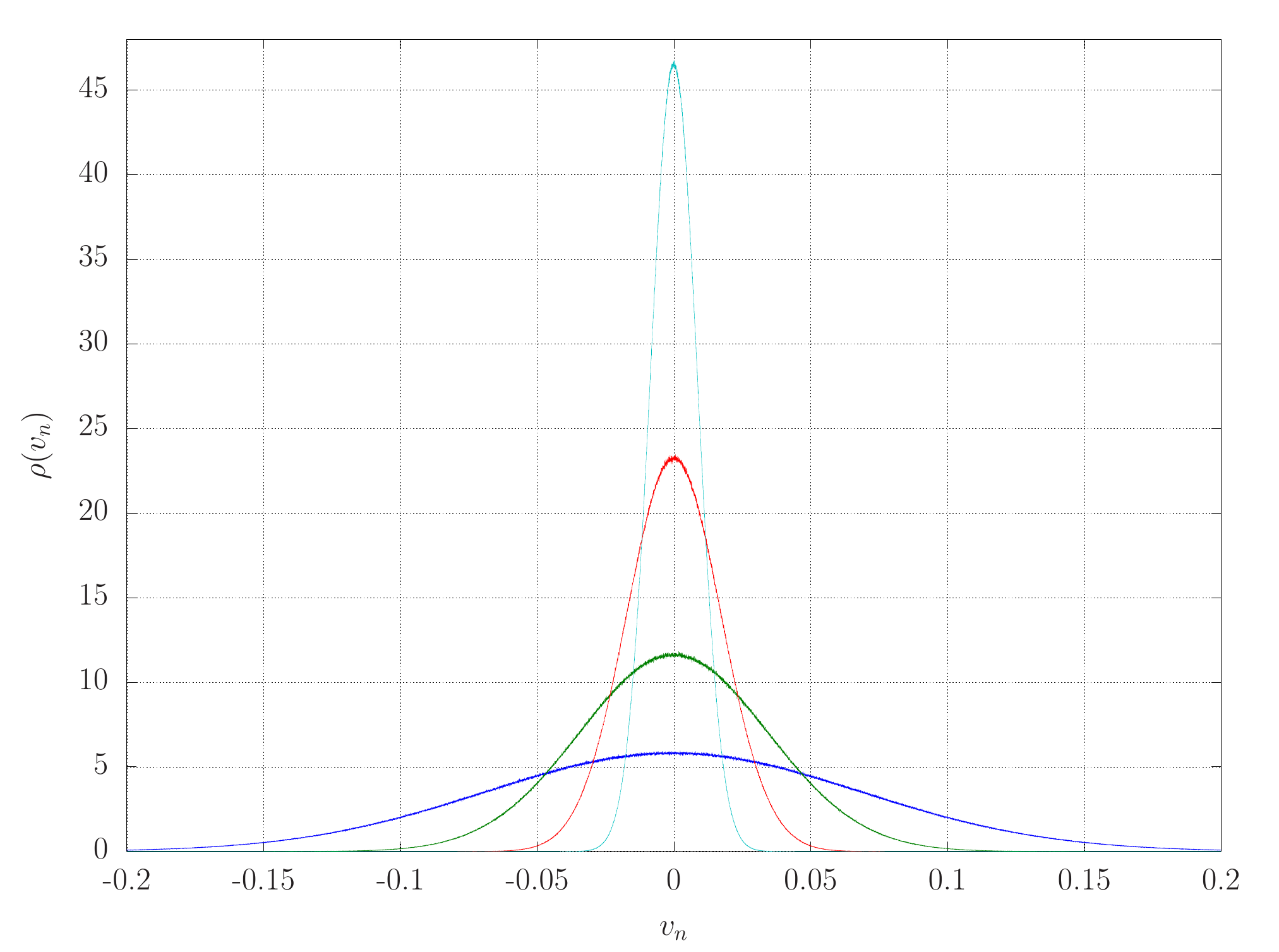}\\
\includegraphics[width=8cm]{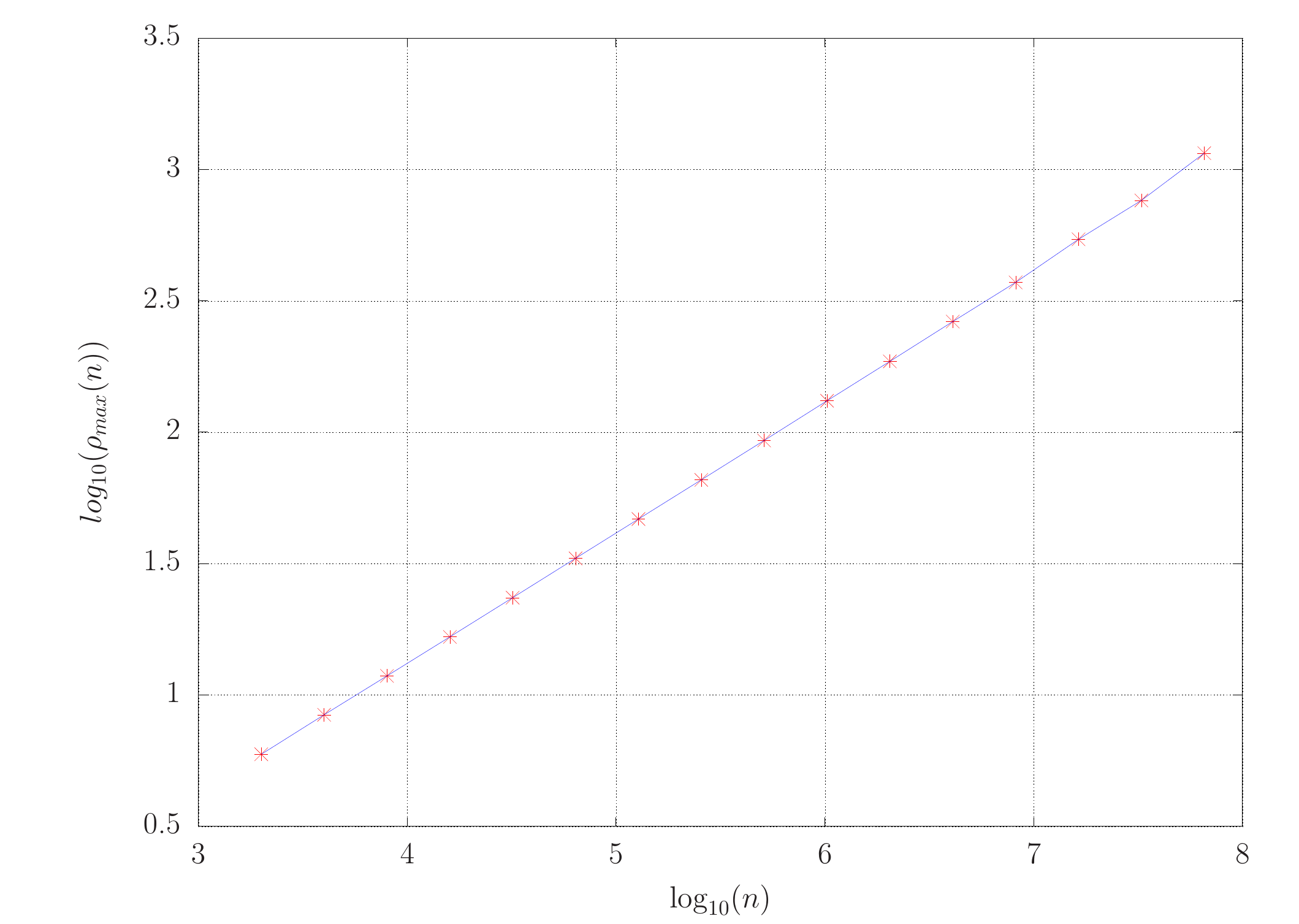}
\par\end{centering}

\protect\caption{$k=10$. Top: distributions of $v_{n}$, for from flattest to thinnest
$n=10^{3},\,8\times 10^{3},\,3.2\times 10^{4},\,1.28\times 10^{5}$. Bottom: evolution
of $\rho_{max}(n)$ versus $n$, in logarithmic scale. One can observe
a scaling $\rho_{max}\sim n^{1/2}$, implying regular diffusive transport
with $\alpha=1/2$. Note that the average has been removed so that distribution are all centered on zero. Data is taken from $1024$ trajectories with initial conditions in the stochastic sea and computed for $10^8$ iterations.\label{fig:distributionsk10}}
\end{figure}
\begin{figure}
\begin{centering}
\includegraphics[width=8cm]{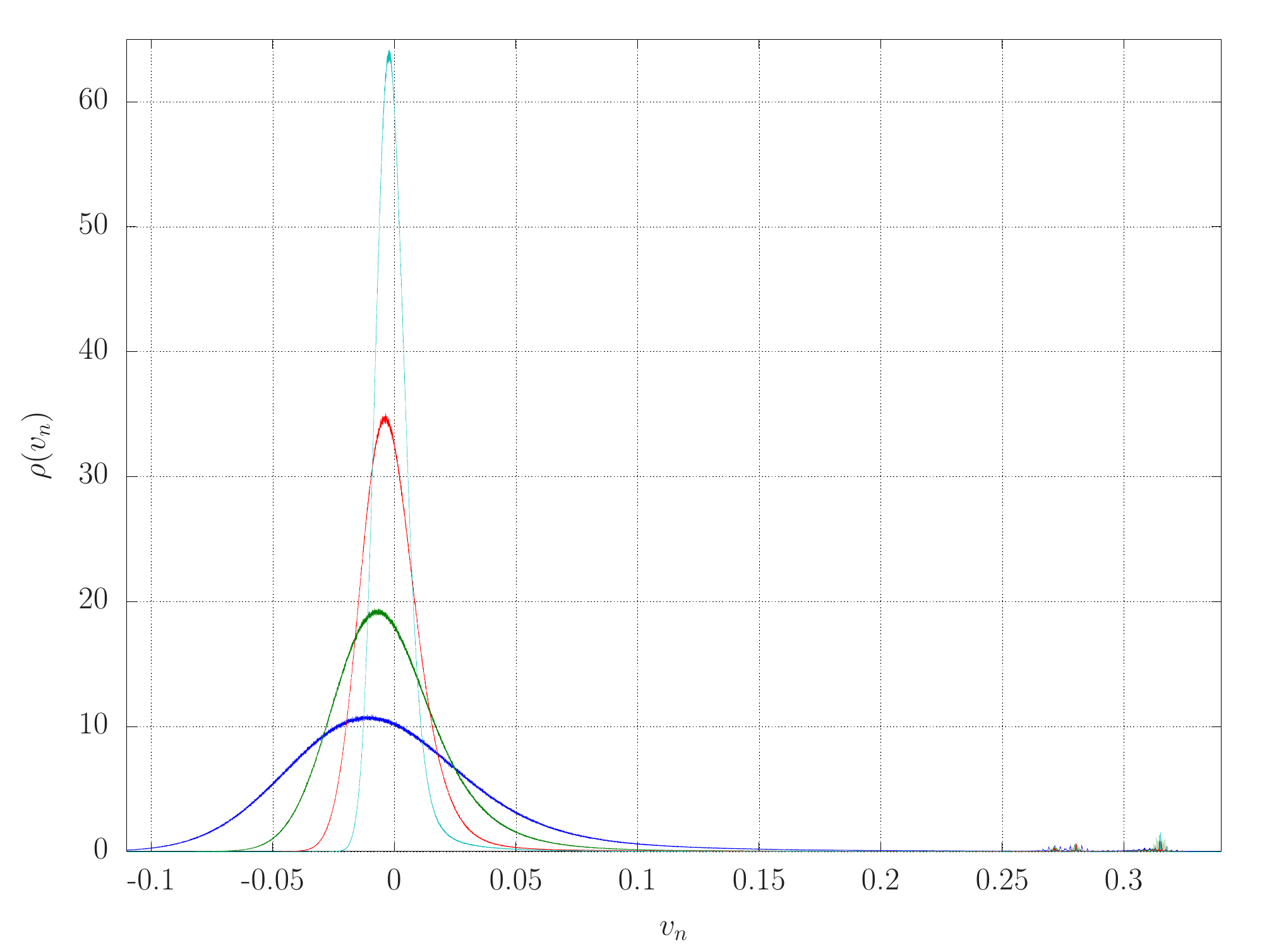}\\
\includegraphics[width=8cm]{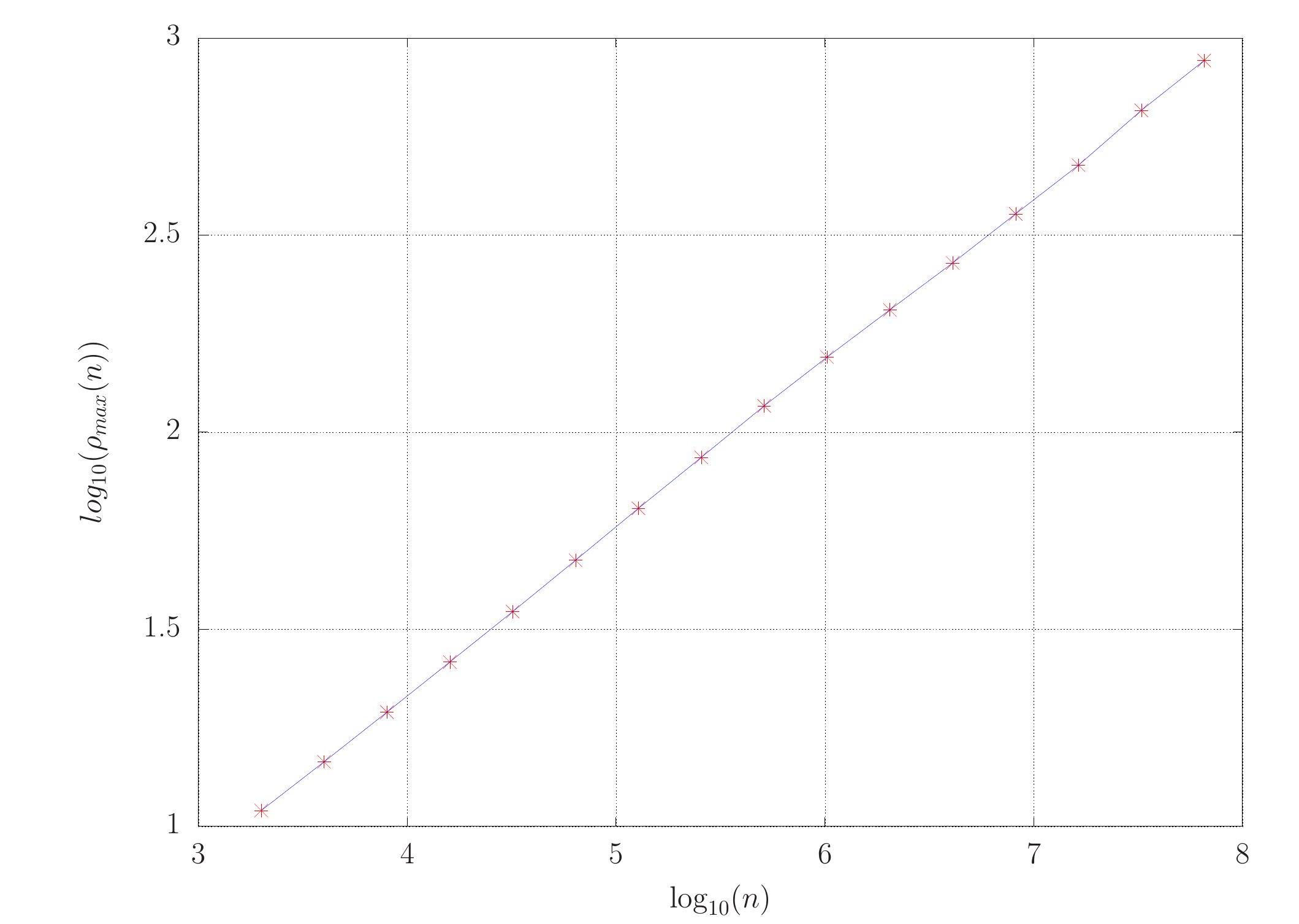}
\par\end{centering}

\protect\caption{$k=2.5$. Top: distributions of $v_{n}$, for from flattest to thinnest
$n=10^{3},\,8\times 10^{3},\,3.2\times 10^{4},\,1.28\times 10^{5}$. Bottom: evolution
of $\rho_{max}(n)$ versus $n$, in logarithmic scale. One can observe
a scaling $\rho_{max}\sim n^{\alpha}$, with $\alpha\approx0.43$
superdiffusive transport. Note that the average has been removed so that distribution are all centered on zero. Data is taken from $1024$ trajectories with initial conditions in the stochastic sea and computed for $10^8$ iterations.\label{fig:distributionsk25}}
\end{figure}
\begin{figure}
\begin{centering}
\includegraphics[width=8cm]{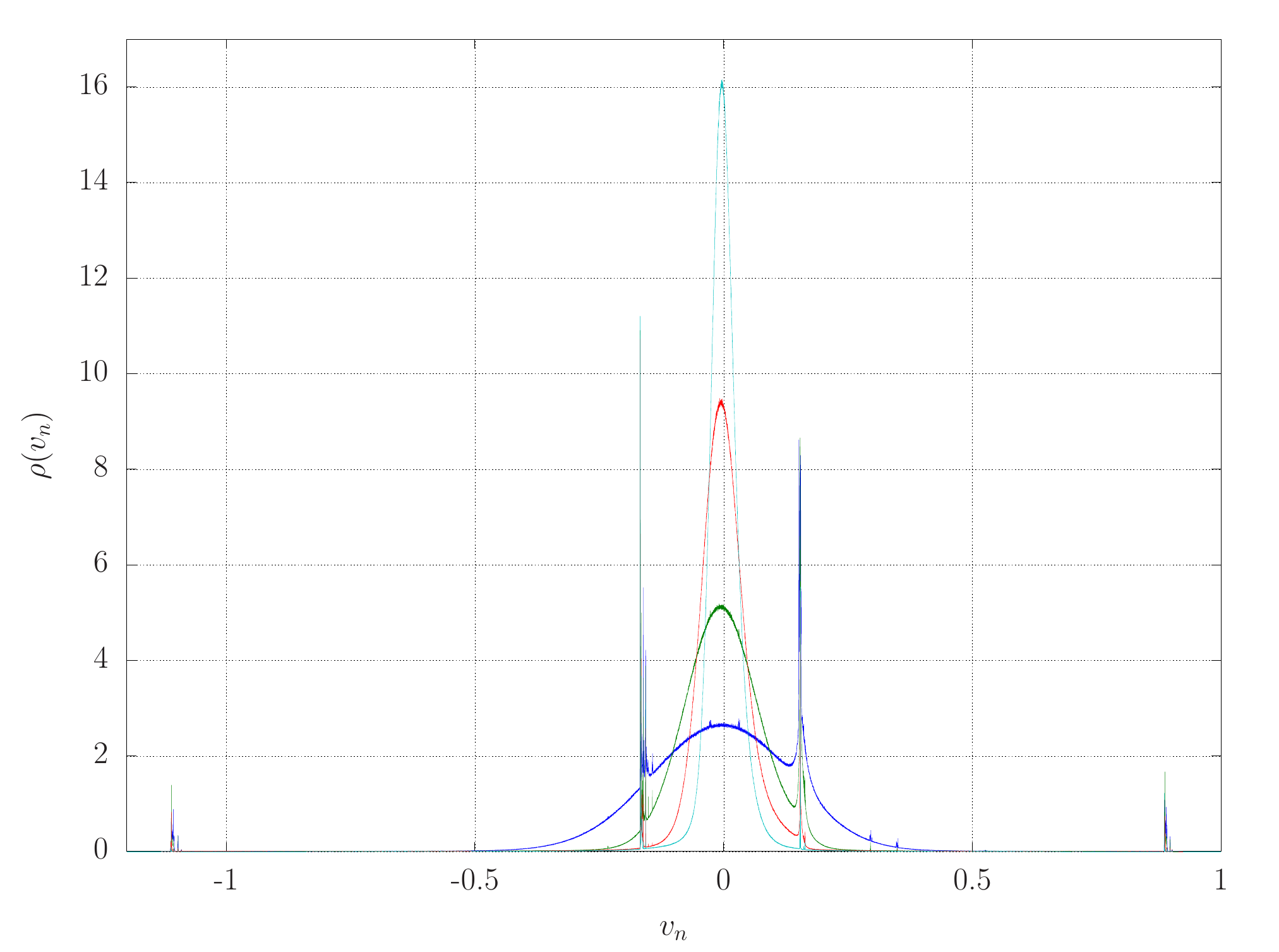}\\
\includegraphics[width=8cm]{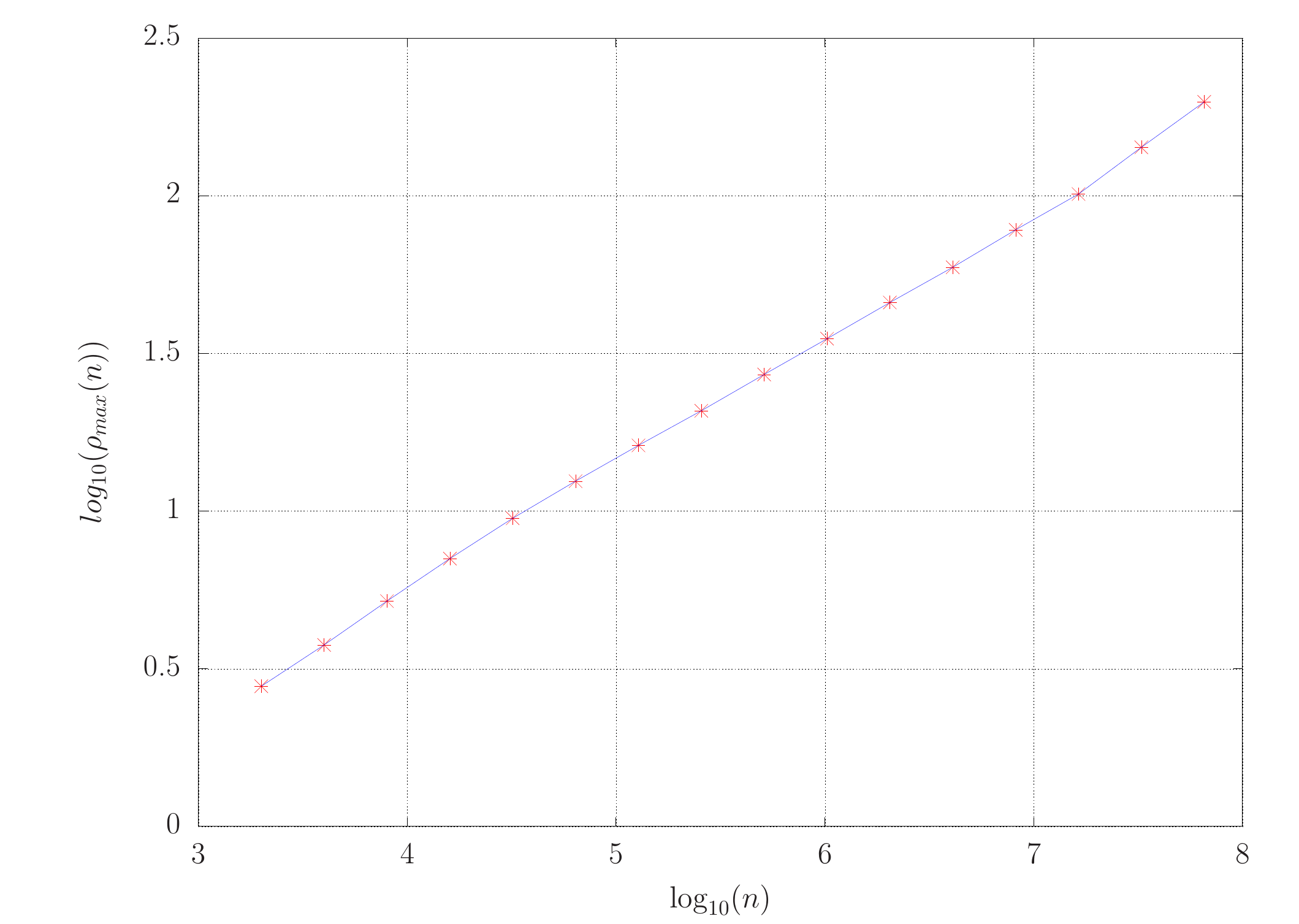}
\par\end{centering}

\protect\caption{$k=1.5$. Top: distributions of $v_{n}$,  from flattest to thinnest distribution
$n=10^{3},\,8\times 10^{3},\,3.2\times 10^{4},\,1.28\times 10^{5}$.  Bottom: evolution
of $\rho_{max}(n)$ versus $n$, in logarithmic scale. One can observe
a scaling $\rho_{max}\sim n^{\alpha}$, with $\alpha\approx0.38$
implying as well superdiffusive transport. Note that the average has been removed so that distribution are all centered on zero. Also there are strong
thin peaks due to stickiness in the distributions, and that $\rho_{max}$
has been computed by measuring the local maximum of the central flat
peak located near $v_{n}=0$. Data is taken from $1024$ trajectories with initial conditions in the stochastic sea and computed for $10^8$ iterations. \label{fig:distributionsk15}}
\end{figure}

We shall now consider the average of $v$ along a typical trajectory.
If the system is ergodic we may naturally expect that the average
will converge to the phase space average of $v$ with the ergodic
measure. In order to assess this statement we shall consider initial
conditions in the stochastic sea, and follow trajectories for large
times; and following the ideas developed in \cite{Leoncini08,LKZ01},
we consider the distribution of finite-time averages. This means that
we shall compute averages of $v$ over finite times, namely for the
given initial condition $(q_i,\:p_i)$ we compute 

\begin{equation}
\bar{v_{i}}(n)=\frac{1}{n}\sum_{k=0}^{n-1}v_{i}(k)\:,\label{eq:speed}
\end{equation}
where $v_{i}(k)$ is the value of the observable
$v$ at the step $k$ for a trajectory given by its initial condition
 $(q_i,\:p_i)$. Let us denote $\rho_{n}(\bar{v})$ the probability density of
the distribution of $\bar{v_{i}}(n)$'s (\ref{eq:speed}). Assuming ergodicity
and we introduce the ergodic average  $\langle v\rangle$. We then have
\begin{equation}
\lim_{n\rightarrow\infty}\rho_{n}(\bar{v})=\delta\left(\bar{v}-\langle v\rangle\right)\:.\label{eq:rho_of_v}
\end{equation}
Since the Dirac function is singular, one simple way to asses the
convergence of $\rho_{n}$, assuming it is relatively smooth, is to
just look at how fast its maximum value $\rho_{max}(n)$ grows towards
$\infty$ with $n$ (see for instance Fig.~\ref{fig:distributionsk10}
top plot). We shall see later that we have to be careful when defining
$\rho_{max}(n)$, but it is easy to figure out that should the dynamics
be sufficiently chaotic, so that a central limit theorem applies,
we can genuinely expect
\begin{equation}
\rho_{max}(n)\sim n^{\alpha}=\sqrt{n}\:,\label{eq:alpha_is_one_half}
\end{equation}
where we defined an exponent $\alpha$ which is equal to $1/2$ in
this case. To convince ourselves we may consider the variables 
\[
y_{i}(n)=\frac{1}{\sqrt{n}}\sum_{k=0}^{n-1}(v_{i}(k)-\langle v\rangle)\:,
\]
whose distribution converges towards a Gaussian when the central limit
theorem applies. Since between $y$'s and $\bar{v}$'s, there is ''just''
a rescaling by $1/\sqrt{n}$, one expects that the mean square displacement
of $\bar{v}$ shrinks as $1/\sqrt{n}$. Then, since the total area
is conserved and equal to one ($\rho$ is a probability density function),
it is natural to infer that $\rho_{n}$ looks more and more like a
Gaussian with a variance that shrinks as $1/\sqrt{n}$ and a maximum
that grows therefore as $\rho_{max}(n)\sim\sqrt{n}$. In fact the
convergence towards the Gaussian depends on the considered point.
Here we  assume that there are no problems with large deviations and that
for large enough $n$ everything is under control (see for instance
\cite{Collet2005,Chazottes99}).

Before moving on to the specific results obtained for the considered
cases, let us emphasize a last feature. For the sake of analogy we previous studies, we shall define what is the arc-length equivalent of the flows for this map as 
\begin{equation}
s_{i}(n)=\sum_{k=0}^{n-1}v_{i}(k)=n\bar{v_{i}}(n)=\sqrt{n}y_{i}(n)+n\langle v\rangle\:.\label{eq:definition of s}
\end{equation}
 It is evidently abusive to denote this as an arclength. Anyhow the origin of anomalous transport is due to the breaking of the central limit theorem. For our case,  this breaking can only occur due to strong time-correlations, i.e memory effects. Since the central limit theorem in dynamical system is applied to a given observable, meaning a function associating a point in phase space to a real number we have to define one. In previous works in low dimensional Hamiltonian flows a suitable choice for an observable appeared to be the norm of the speed in phase space. Using this observable attached to no particular coordinate system ended up as a good choice, especially when the phase space was bounded. When dealing with the standard map, the notion of speed per se has no meaning, so for the sake of analogy with previous work we here choose a specific observable $v$ which ``resembles'' a speed, and its associated displacement $s$. It is important to remind the reader that, except of a few specific observables, the nature of transport will not depend on the choice of the observable.

We now consider the transport properties using $s$. If the central limit theorem applies, one will expect that the varianceof the distribution of $s_{i}$'s grows like $\sqrt{n}$, so that
the second moment has a characteristic exponent $\mu=1$.

Let us now envision a situation for which transport is anomalous,
for instance super-diffusive, with so called fat tails like power
law decreasing ones for instance giving rise to a transport exponent
$1<\mu<2$. The scaling relation (\ref{eq:definition of s}) between
$s$ and $\bar{v}$ still holds, which means that the variance of
$\bar{v}$ decreases not as fast as for the Gaussian case, since we
still have area conservation, we can expect that the maximum grows
slower than $\sqrt{n}$ (if the distribution is flat enough at the
top) and thus that the characteristic exponent related to the growth
of the maximum of the distribution $\alpha$ is such that $\alpha<1/2$.
In fact since, we will assume that these power law behaviors of the
maximum and the variance are valid. Then since the total probability
is conserved and equal to one for each time $n$, a fact that can
be viewed as area conservation under the drawing of the function $\rho_{n}(\bar{v})$.
Since we are dealing with scaling laws, we can make the rather crude
approximation that the probability density function $\rho_{n}(\bar{v})$
can be approximated by a rectangle function. Then considering the
conservation of the area $S=\rho_{max}(n)\times\sqrt{\sigma(n)}$,
where $\sigma(n)$ is the variance of the $\bar{v_{i}(n)}$, we directly
obtain a relation between $\mu$ and $\alpha$ (see \cite{Leoncini08}
) which writes
\begin{equation}
\alpha=1-\frac{\mu}{2}\:.\label{eq:beta_and_alpha}
\end{equation}
In any case even if Eq.(\ref{eq:beta_and_alpha}) does not hold, the
previous considerations offer a ''different'' possible way to characterize
anomalous transport depending on the values of $\alpha$:
\begin{itemize}
\item If $\alpha\ne1/2$ transport is anomalous, moreover

\begin{itemize}
\item if $\alpha>1/2$ we expect sub-diffusive transport
\item if $\alpha<1/2$ we expect superdiffusive transport.
\end{itemize}
\end{itemize}
In order to confirm these statements for the considered system (\ref{eq:Standard map}),
we as well compute its transport properties. In order to characterize
transport, we follow  \cite{Castiglione99}
and consider the transport properties related to the chosen observable,
meaning we compute the different moments of (\ref{eq:definition of s}),
from which we extract a characteristic exponent 
\begin{equation}
M_{q}(n)=\langle|s_{i}(n)-\langle s_{i}(n)\rangle|^{q}\rangle\sim t^{\mu(q)}.\label{eq:Moments_def_mu(q)}
\end{equation}
\begin{figure}
\begin{centering}
\includegraphics[width=8cm]{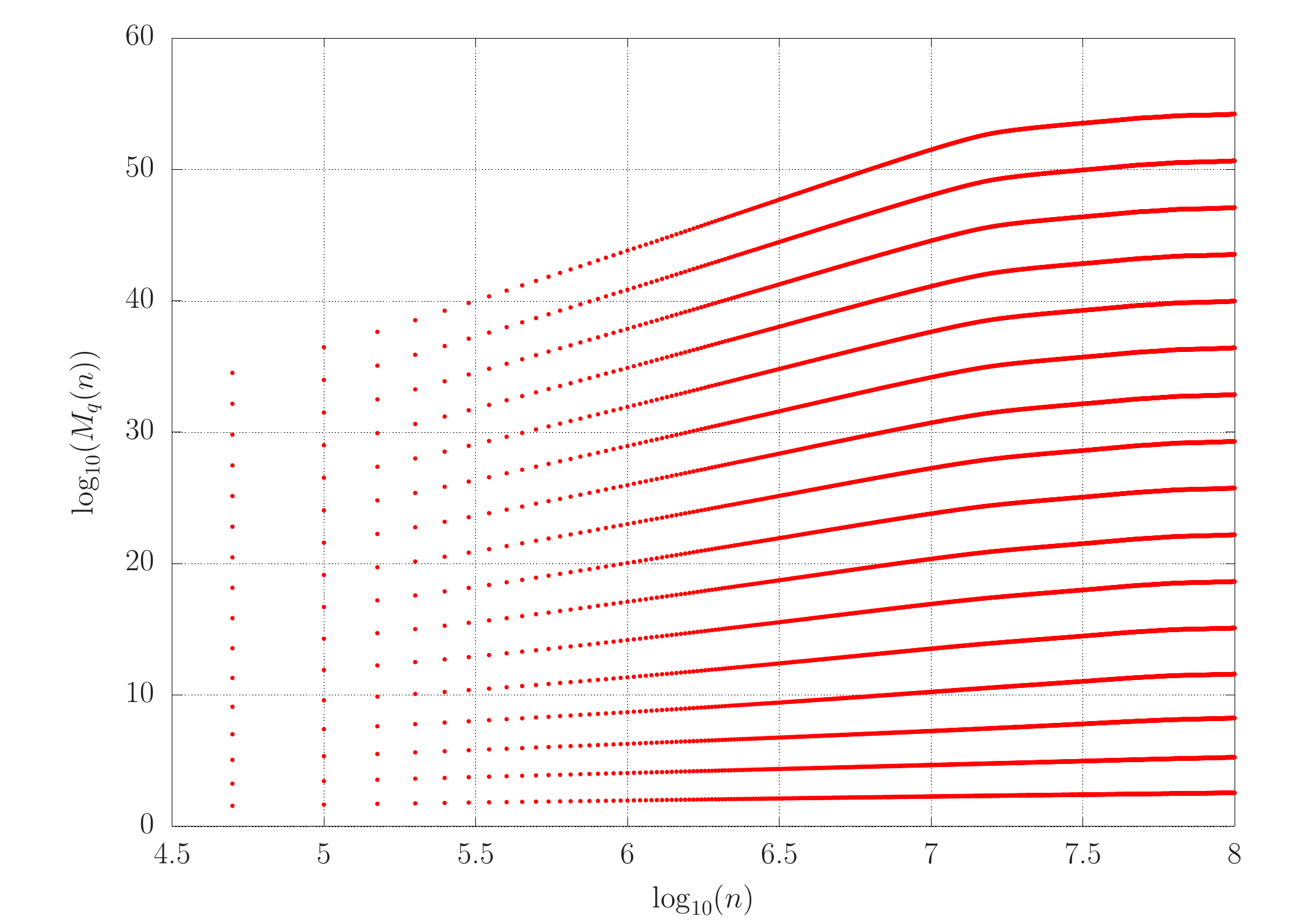}\\
\includegraphics[width=8cm]{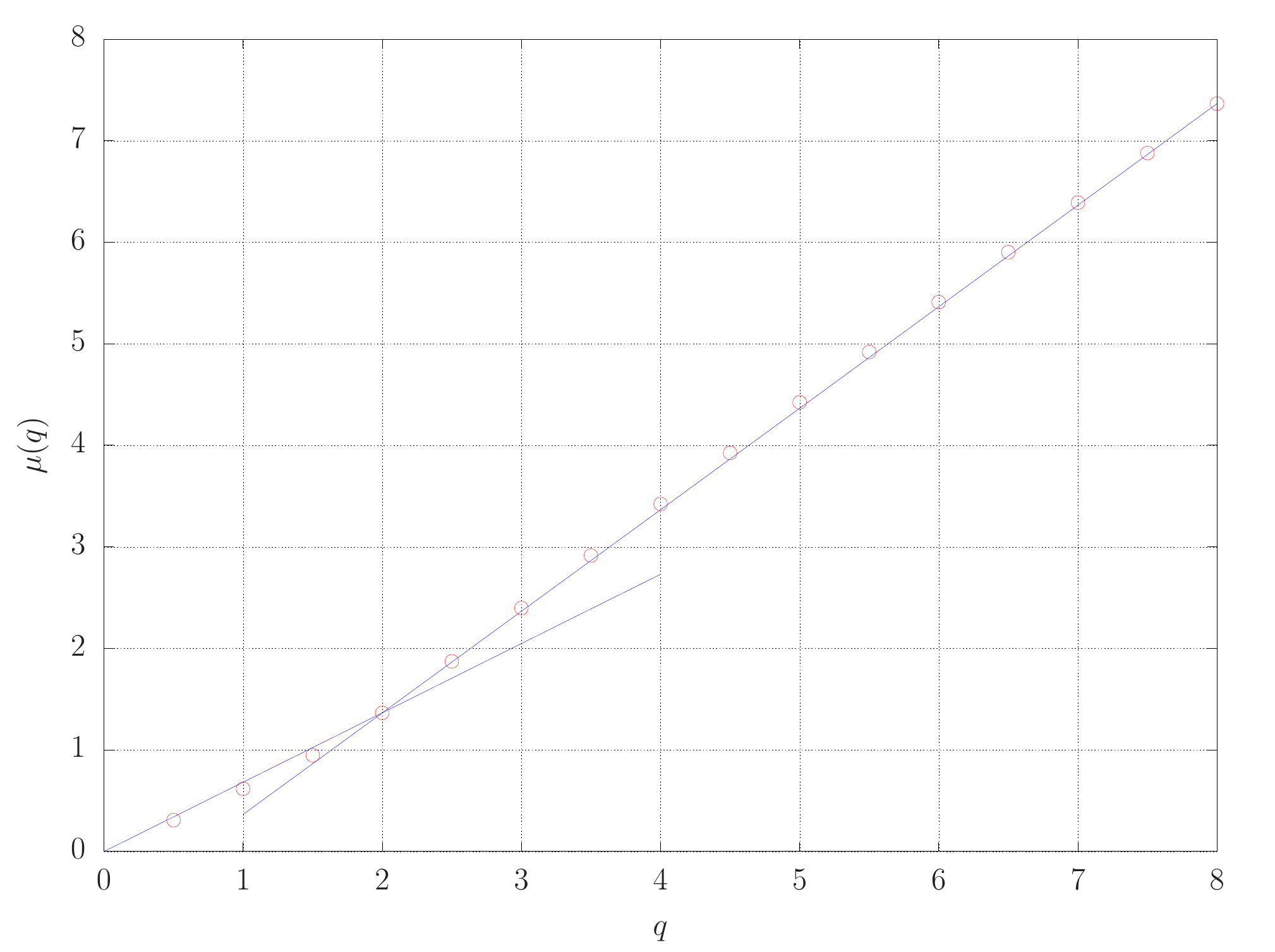}
\par\end{centering}

\protect\caption{$k=1.5$. Top: moments versus time of $s$. Bottom: characteristic
exponent versus moment order. These exponent have been computed using
the linear portion of the top picture. This nonlinear behavior is
typical of strong anomalous transport. We have super-diffusive transport
and $\mu(2)\approx1.3$.\label{fig:momentsk15}}
\end{figure}
\begin{figure}
\begin{centering}
\includegraphics[width=8cm]{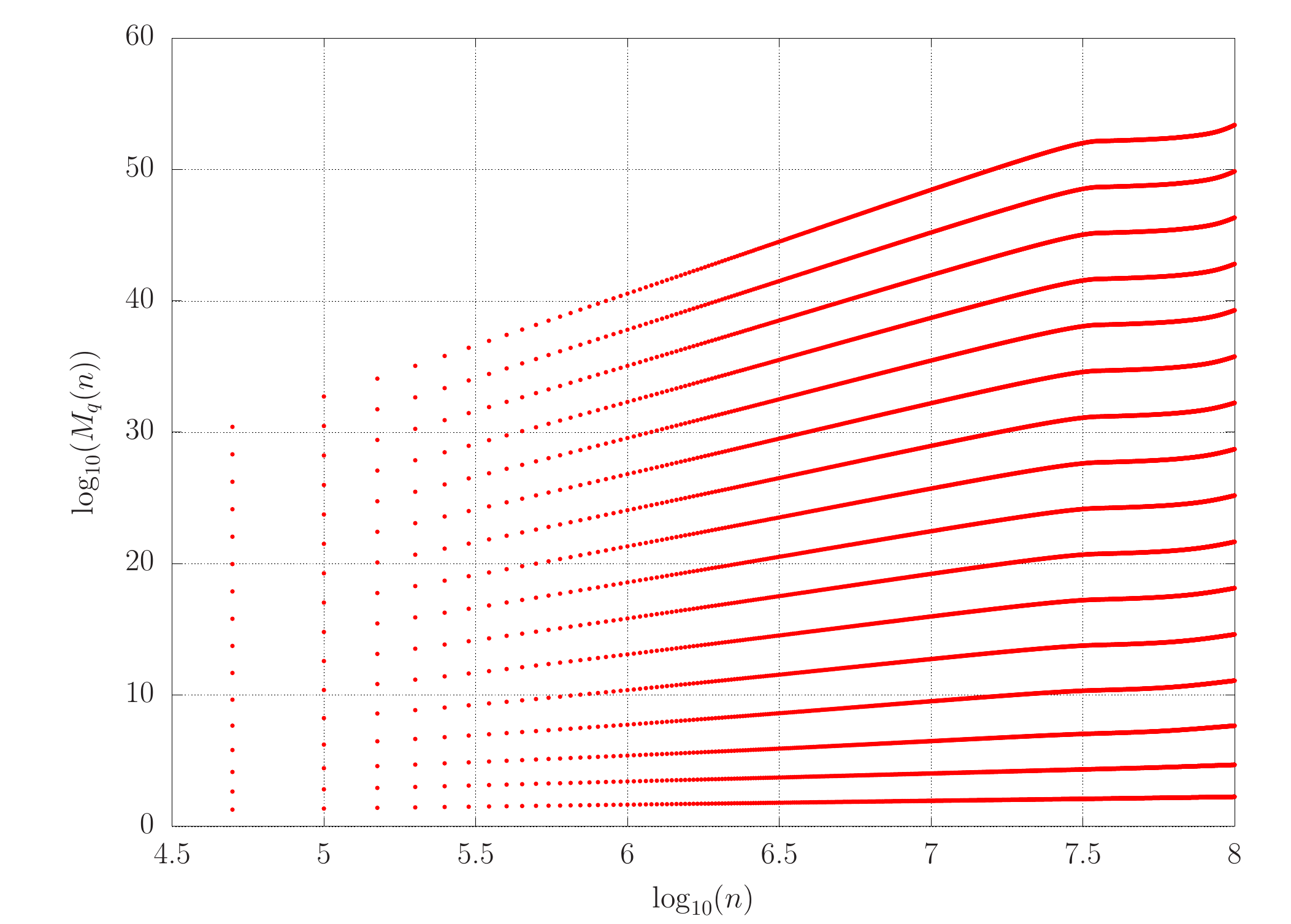}\\
\includegraphics[width=8cm]{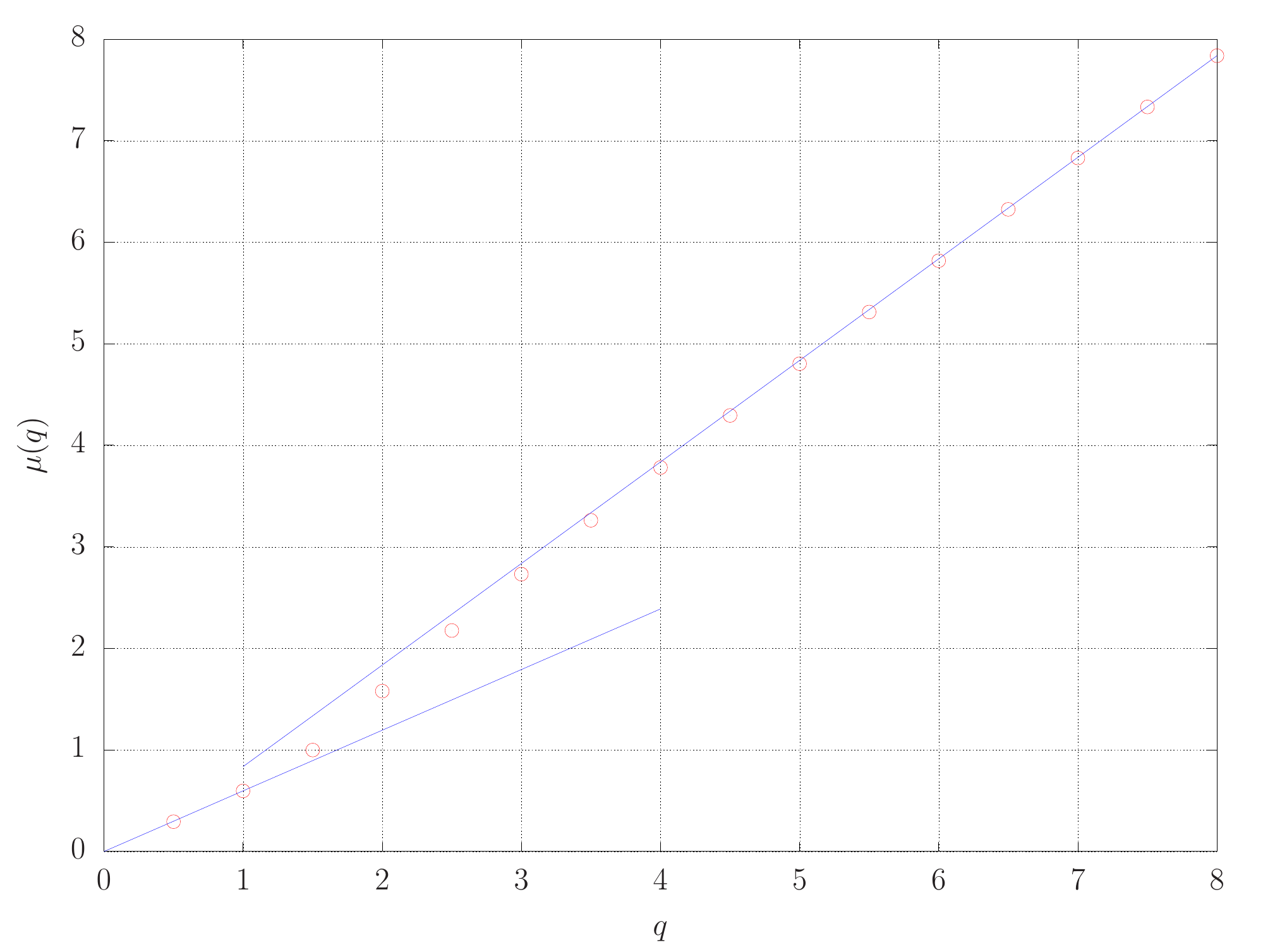}
\par\end{centering}

\protect\caption{$k=2.5$. Top moments versus time of $s$. Bottom characteristic
exponent versus moment order. These exponent have been computed using
the linear portion of the top picture. This nonlinear behavior is
typical of strong anomalous transport. We have super-diffusive transport
and $\mu(2)\approx1.67$.\label{fig:momentsk25}}
\end{figure}
\begin{figure}
\begin{centering}
\includegraphics[width=8cm]{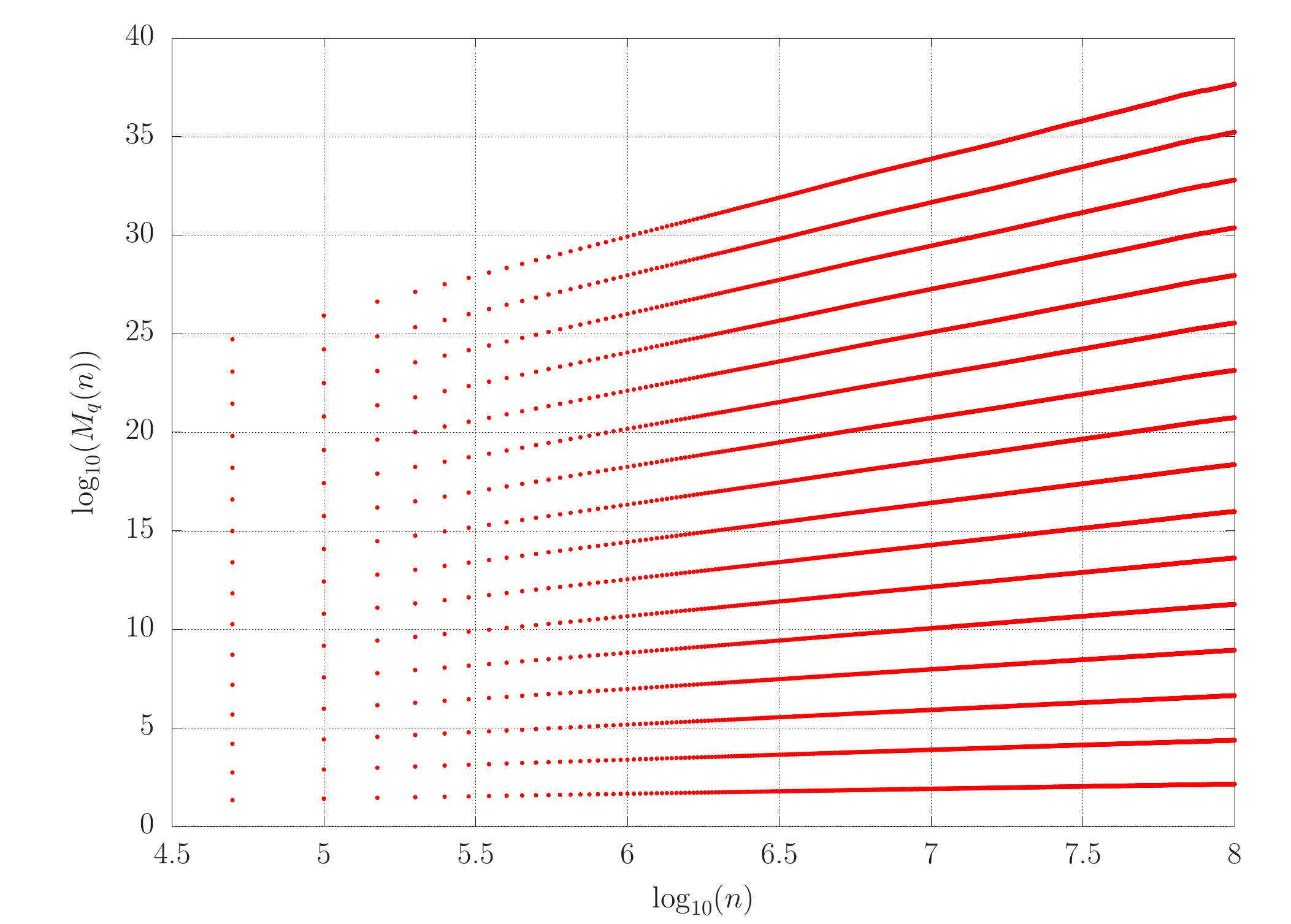}\\
\includegraphics[width=8cm]{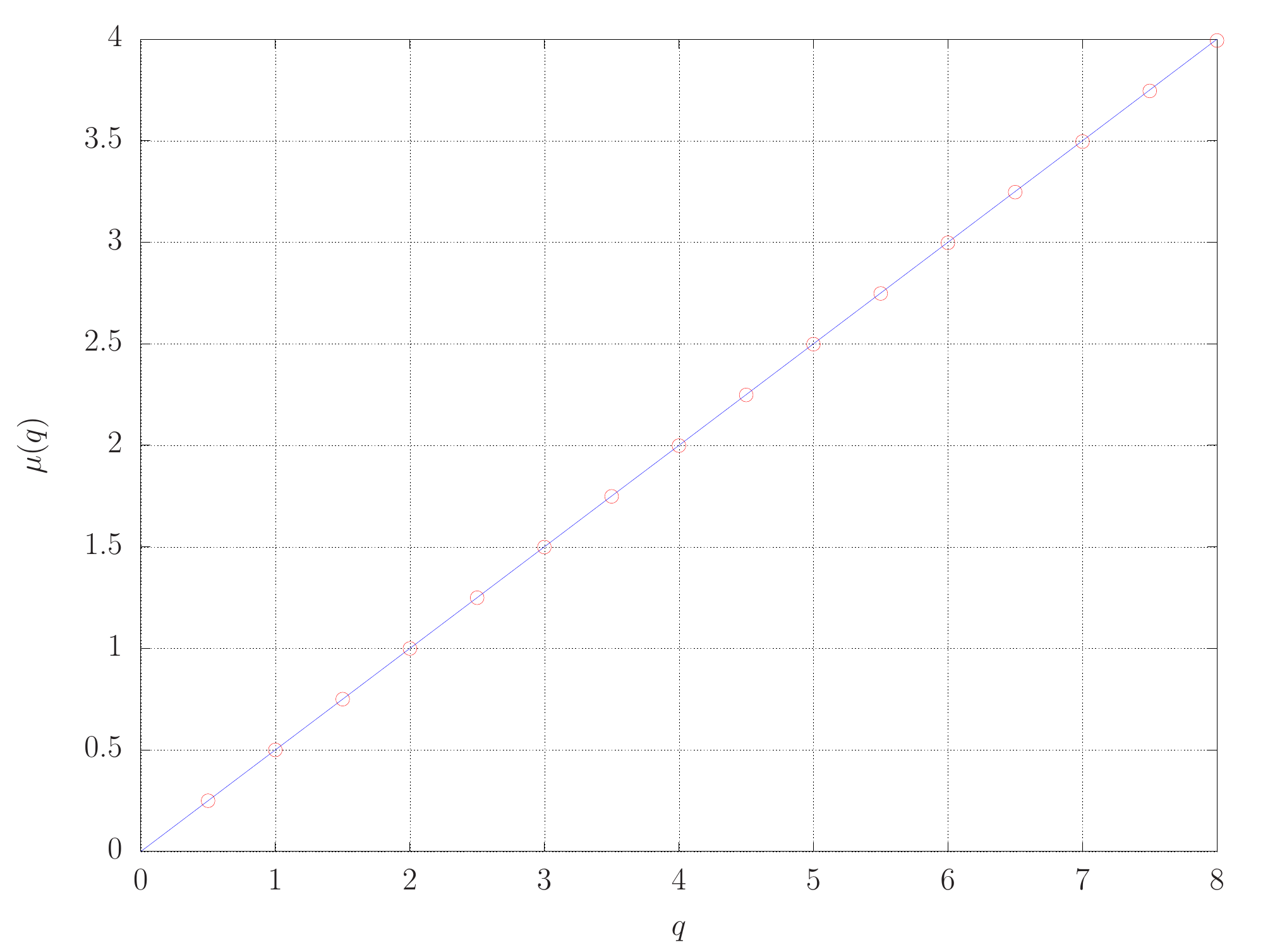}
\par\end{centering}

\protect\caption{$k=10$. Top: moments versus time of $s$. Bottom: characteristic
exponent versus moment order. We observe the linear behavior with
slope $1/2$ characteristic of diffusive transport.\label{fig:momentsk10}}
\end{figure}
 The second order exponent $\mu=\mu(2)$ characterizes the anomalous
and diffusive transport. We recall that transport is so-called diffusive
if $\mu(2)=1$ , and in all other cases the transport is anomalous.
More specifically, if $\mu(2)<1$ the transport is sub-diffusive and
if $\mu(2)>1$, it is super-diffusive.

In order to check all these assumptions we considered the standard
map dynamics in the three specific case described in Figs.~\ref{fig:Phase-space-visualsationk=00003D1.5},\ref{fig:Phase-space-visualsationk=00003D2.5},\ref{fig:Phase-space-visualsationk=00003D10}.
We numerically computed the evolution of $1024$ different
trajectories, for which we computed $10^{8}$ steps. Initial conditions where taken in a square of side 0.01 centered on the point $(p,\:q)=(0.1,\:0.1)$. Records of $s$'s
and positions are taken every $10^{3}$ steps, meaning that we have
about $10^{8}$ records to compute our distribution and moments. When
computing histograms to obtain the distribution, we sampled equally
spaced $5\,10^{4}$ bins between the minimal and maximal value of
the $\rho(v)$'s. We insist on the fact that when binning is not adequate
we may end up for large times having all the $\bar{v_{i}}(n)$ accumulating
in the same interval, when this happens the distribution we obtain
is effectively a rectangle function, whose maximum does not grow,
a phenomenon that can be monitored using $\rho_{max}(n)$. We may
also end up having problems with the accuracy of our data due to the
finite precision of our real numbers (we considered here double precision
numbers with 16 characteristic digits), which would lead to a similar
phenomenon but may as well accelerate the growth of $\rho_{max}(n)$.
Finally if our sample is to refined in regards to our amount of data,
the distribution becomes too discontinuous to mean anything and we
can not rely on our analysis. Given these constraints, we settled
for the aforementioned numbers, and checked the stability of the results
versus different binning strategies. Increasing the duration of the
trajectories (computing more time steps) will lead to problems of
accuracy due to finite precision, and it would thus necessitate the
use of computing trajectories with higher precision. The other strategy
would be to compute more trajectories, i.e take more initial conditions,
however this leads to a bias because it would only work assuming that
the ergodic measure in the stochastic sea is related to the Lebesgue
one, and to avoid this bias we prefer to consider finite time portions
of trajectories computed over large time. Of course using more powerful
computers, computing more data is always possible, but the presented results
are sufficiently accurate in order to confirm our conclusion on the nature of transport
and to link between the exponents.

We now return to the analysis of our results.

The first simpler case is the fully chaotic one, namely the case for
which $K=10$ (Fig.~\ref{fig:Phase-space-visualsationk=00003D10}).
We first represented in Fig.~\ref{fig:distributionsk10}, the distributions
resulting from different time averages, and the evolution of the maximum
$\rho_{max}$ versus the length over which we average $n$. We find
as anticipated that $\alpha=1/2$ and thus expect to have normal diffusive
transport. We then move on to the two other cases with mixed phase
space, meaning with a non-unique ergodic measures. For these, we consider
initial conditions in the so-called stochastic sea. Should we consider
trajectories inside the regular regions they would have a ballistic
like contribution, and would not belong to the same ergodic component.
For the situation displayed in Fig.~\ref{fig:distributionsk25},
we can notice in the distribution that small peaks (bumps) appear near $v_{n}\approx0.3$,
these peaks correspond to the stickiness phenomenon and portions of
trajectories that remain a long time around a regular island, giving
rise to memory effect and Lévi flights (see for instance \cite{LKZ03}).
The presence of these peaks naturally affects $\alpha$, and we find
$\alpha\approx0.43$, meaning we expect super-diffusive transport.
For the last considered case $K=1.5$ displayed in Fig.~\ref{fig:distributionsk15},
we observe a similar phenomenon, the peaks are however much sharper
than in the $K=2.5$ situation. In fact for small values of $n$ the
presence of these peaks affects the value of $\rho_{max}(n)$, in
order to circumvent this problem we considered in Fig.~\ref{fig:distributionsk15}
only the local maximum centered around the final average value; for
large values of $n$ this does not affect the value of $\rho_{max}(n)$,
but it does for small $n$'s. In this case, the polynomial behavior
appears to be only roughly correct, and we measure $\alpha\approx0.38$,
so we expect as well super-diffusive transport. If we have some faith
in Eq.(\ref{eq:beta_and_alpha}), we expect as well a larger second
characteristic exponent for transport than for the $K=2.5$ case. 

In order to verify our conclusions we computed the transport properties
from the different data sets. Meaning we computed the different moments
$M_{q}(n)$ and extracted from each a typical characteristic exponent
$\mu(q)$. The results are displayed in Figs.~\ref{fig:momentsk15},
\ref{fig:momentsk25} and \ref{fig:momentsk10}. 

We shall discuss these results starting from the simplest case, namely
the fully chaotic situation corresponding to $K=10.$ The results
of this case is represented on Fig.~\ref{fig:momentsk10}, as expected
in this testbed situation we recover the features of Gaussian diffusive
transport, meaning a second moment $\mu(2)=1$, and a linear behavior
of the moments $\mu(q)=q/2$ \cite{Castiglione99}. Given the previous
results and the value of $\alpha=1/2$ measured previously, we have
a first setting for which the equation Eq.(\ref{eq:beta_and_alpha})
holds. Considering a more complicated case, namely when $K=1.5$ (Fig.~\ref{fig:momentsk15}),
the nonlinear behavior of the function $\mu(q)$ with $\mu\sim q$
for large values of $q$. According to the definitions given in \cite{Castiglione99},
this behavior implies that we are facing multi-fractal transport or
so called strong anomalous transport. Anyhow, we still find a good
agreement between the measured value of the second moment $\mu(2)\approx1.3$
and what the value of we computed as well $\alpha=0.38$, the equation
Eq.(\ref{eq:beta_and_alpha}) still holds and we insist as well that
since $\alpha<1/2$ we found as expected super-diffusive transport.

It is in fact for the intermediate case $K=2.5$, with the small peaks
that we got unexpected results. Indeed, in Fig.~\ref{fig:momentsk25},
we can notice that the second moment exponent is located in the area
corresponding to the change of asymptote between two linear regimes.
The measured value of the transport exponent gives $\mu(2)\approx1.67$.
This measurement allows to clearly state that the equation Eq.(\ref{eq:beta_and_alpha})
is not general enough and does not hold for this particular situation.
Indeed we measured $\alpha\approx0.43$, while should Eq.(\ref{eq:beta_and_alpha})
be true we would get $\alpha_{expected}=0.17$. Note though that the
measured value of $\alpha$ correctly concludes on the nature of the
transport which is also super-diffusive. 

Measuring this exponent is therefore a good indicator of the nature
of transport, but how is it related to the measured transport exponent.
In fact in all of our previous computations, as well as in the results
presented in \cite{Leoncini08}, Eq.(\ref{eq:beta_and_alpha}) was
holding. In order to explain what is going on we shall have a closer
look at the exponent versus moment order figures displayed in Figs.~\ref{fig:momentsk15},
\ref{fig:momentsk25} and \ref{fig:momentsk10}. We can notice that
both cases for which the expression Eq.(\ref{eq:beta_and_alpha})
holds, correspond to an exponent whose value can be obtained by simply
extrapolating the linear behavior of the function $\mu(q)$ for small
values of $q$, i.e the small moments linear behavior. This feature
was also true in the results presented in \cite{Leoncini08}, where
the expression Eq.(\ref{eq:beta_and_alpha}) was first derived. A
natural simple conclusion is then to conclude that the same is true
for this $K=2.5$ problematic case. In order to check this we perform
the same extrapolation from the low values of $q$ to $q=2$ of $\mu(q)$
for the case $K=2.5$ in Fig.~\ref{fig:momentsk25}. We observe as
well a linear behavior in Fig.~\ref{fig:momentsk25} for low moments
which we can extrapolate, we then measure $\mu_{extrapolated}(2)\approx1.2$.
This extrapolated value is in remarkable agreement with what we would
expect using formula (\ref{eq:beta_and_alpha}), indeed we would
then obtain $\alpha=0.4$ close to the measured $\alpha=0.43$ one. 

From these last remarks and measurement we may decide that this linear
extrapolation is actually a more general expression relating the exponent
$\alpha$ to the transport moments $\mu(q)$. We may therefore speculate
that a correct expression is

\begin{equation}
\frac{d\mu}{dq}(0)=1-\alpha\:,\label{eq:beta_and_alpha-1}
\end{equation}
meaning that the exponent $\alpha$ is directly related to the slope
at the origin of the function $\mu(q)$. We conjecture, that in all situations
for which the previous expression Eq.(\ref{eq:beta_and_alpha}) holds,
the observed value $\mu(2)$ can be extrapolated from the slope at
the origin. Eq.~(\ref{eq:beta_and_alpha-1}) provides therefore a
more general setting that is compatible with previous results, but
also when facing situations like the one for $K=2.5$.

In order to understand why actually this Eq.~(\ref{eq:beta_and_alpha-1})
is probably more sound, we are looking again at the moments displayed
in Figs~\ref{fig:momentsk25}~and~\ref{fig:momentsk15}. In these
pictures we notice that the evolution is not of the power law type
for large times, with some not standard more or less flat evolution.
In fact when looking at the data, this behavior has a simple explanation.
In fact this may be due to the finite time sampling and finite number
of trajectories. Indeed let us consider the situation of a trajectory
captured in a long Lévy flight (captured in a sticky set for instance)
which suddenly stops ''flying'' (goes back to the stochastic sea).
When computing high order moments, we can naturally state that these
will be essentially dominated by this ballistic excursion, and when
the flight stops, the moments will then remain more or less constant
until the rest of the trajectories finally manage to reach such large
excursion. On the other side this effect is less visible on lower
moments. Indeed the inversion of convexity for $q<1$ puts less emphasis
on large excursions. The same is true for the exponent $\alpha$.
Indeed the portion of a trajectory caught in a flight will contribute
to a specific peak different than the bulk one (see for instance Fig.~\ref{fig:distributionsk15}),
then when the flight stops, it will take a while before the time average
of the speed of this portion of trajectory enters the bulk, and starts
contributing to the main peak and thus influence the growth of its
tip, i.e the value of $\alpha$.

\section{Conclusion}

In this paper we  studied transport properties using distributions
of finite time observable averages and monitoring the time evolution
of its maximum in the spirit of the work presented in \cite{Leoncini08}.
In order to have access to large amount of data, we opted to reduce
the numerical computation of the flow, while still considering a Hamiltonian
system; we opted for the kicked rotor system which can be reduced
to the standard map, which we settled for as a test case study. In
this setting we have shown that $\alpha$, the characteristic exponent
of the time evolution of the maximum of finite time averaged observable 
distributions, conveys good information relative to the nature of
transport. Indeed, we confirmed the nature of transport with respect
to the values of $\alpha$, and we found super-diffusive transport
when $\alpha<1/2$ which corresponds to situations of a mixed phase
space, while we recovered Gaussian transport for the fully chaotic
regime and $\alpha=1/2$. We as well confirmed the multi-fractal nature
of transport in the standard map for the cases with mixed phase space
already discussed in \cite{Bouchara2012} and reference therein. Finally
we propose a link between the characteristic exponent of transport
moments and $\alpha$, which generalizes the results obtained in \cite{Leoncini08}.
This expression appears to encompass as well situations when the exponent
of the second moment lies in the nonlinear zone of the curve relating
the characteristic moment exponent $\mu(q)$ versus the order of the
moment $q$. Using this new relation, we find a good agreement between
the measured values of $\alpha$ and the behavior of the characteristic
transport exponent for low order. Finally one could ask the reason
on why to introduce this $\alpha$ exponent. Besides its intrinsic
interest we found in the end that it is computationally easier and
faster to compute. Indeed, as long as good care of histogram computation
and bin size is performed, the linear scaling law to extract its value
appears to be much easier than when dealing with transport for which
this is not always obvious especially when transport is super-diffusive.

\section*{Acknowledgments}
This work has been carried out thanks to the support of the A*MIDEX project (n° ANR-11-IDEX-0001-02) funded by the  ``Investissements d'Avenir'' French Government program, managed by the French National Research Agency (ANR), SV and XL were also supported by the project MOD TER COM of the french
Region PACA, SV acknowledges also support from the ANR grant PERTURBATIONS.

\bibliographystyle{elsarticle-num}

\end{document}